\documentclass{jpsj3}
\usepackage{graphicx}

\begin{document}
\title{Nonequilibrium work on spin glasses in longitudinal and transverse fields}
\author{Masayuki Ohzeki, Hitoshi Katsuda$^1$ and Hidetoshi Nishimori$^1$}

\inst{
Department of Systems Science, Graduate School of Informatics, Kyoto University, Yoshida-Honmachi, Sakyo-ku, Kyoto 606-8501, Japan\\
$^1$Department of Physics, Tokyo Institute of Technology, Oh-okayama, Meguro-ku,
Tokyo 152-8551, Japan} 
\abst{
We derive a number of exact relations between equilibrium and nonequilibrium quantities for spin glasses in external fields using the Jarzynski equality and gauge symmetry.
For randomly-distributed longitudinal fields, a lower bound is established for the work done on the system in nonequilibrium processes, and identities are proven to relate equilibrium and nonequilibrium quantities.
In the case of uniform transverse fields, identities are proven between physical quantities and exponentiated work done to the system at different parts of the phase diagram with the context of quantum annealing in mind.
Additional relations are given, which relate the exponentiated work in quantum and simulated (classical) annealing.
It is also suggested that the Jarzynski equality may serve as a guide to develop a method to perform quantum annealing under non-adiabatic conditions.
}

\kword{quantum annealing, Jarzynski equality, spin glass, gauge symmetry}
\maketitle

\section{Introduction}
The remarkable recent developments in nonequilibrium statistical physics, the Jarzynski equality \cite{J1,J2} and fluctuation theorem \cite{J3,J4,J5}, have given a paradigm to understand dynamical behavior through a direct relationship between a nonequilibrium process and equilibrium states (Jarzynski equality) or the probabilities of a nonequilibrium process and its inverse process (fluctuation theorem). 
The purpose of the present paper is to examine dynamical properties of spin glasses by use of the Jarzynski equality and the fluctuation theorem and establish exact identities and inequalities involving equilibrium and nonequilibrium quantities for spin glasses.

Properties of spin glasses have been studied extensively for many years by experimental, numerical, and analytical methods \cite{Rev1,Rev2,Rev3}.
Although most of the theoretical aspects have been understood fairly satisfactorily at the mean-field level \cite{MPV}, it is still difficult to establish analytical results for finite-dimensional systems. 
One of the exceptional successful approaches is an analysis by the gauge symmetry \cite{HN81,HNbook}. 
Not only equilibrium quantities but also nonequilibrium behavior can be evaluated without approximations \cite{Ozeki}. 
Our previous study \cite{ON} has revealed several new types of exact equalities between nonequilibrium processes and equilibrium properties of spin glasses by use of the gauge symmetry in addition to the above-mentioned nonequilibrium relations. 
In that paper, we have considered an application of the method of gauge transformation to the Jarzynski equality to shed new light on the possibilities to use a relatively unconventional sampling method, annealed importance sampling or its improvement known as the population annealing \cite{NJ,Iba,Pop1}, in the theoretical studies of spin glasses.
These methods use a relation analogous to the Jarzynski equality while changing the temperature similarly to simulated annealing \cite{SA} and show outstanding performance comparable to the exchange Monte Carlo method \cite{EX}. 
However, our previous study treated the change in temperature and fell short of the investigation of the performed work corresponding to the direct change of the parameters in the Hamiltonian.
In the present paper, we revisit the problem of nonequilibrium processes in spin glasses and develop a theory to analyze the work done to the system by the change of the strengths of external fields.
The resulting identities and inequalities represent simple relations between equilibrium and nonequilibrium quantities in spin glasses in external fields, which should be added to the list of rare exact results on static and dynamic properties of spin glasses

This paper is organized as follows. 
In \S 2, we analyze the work performed on spin glasses by randomly-distributed longitudinal fields. 
We then discuss the possibility of non-adiabatic quantum annealing starting from equilibrium ensembles and analyze the work performed on spin glasses by the transverse field. 
In addition, we give highly non-trivial relations between two completely different processes, simulated and quantum annealing. 
The last section will summarize this paper.

\section{Spin glass in random fields with gauge invariance}
\subsection{Model}
In the present section we discuss the $\pm J$ Ising model of spin glasses in random fields on an arbitrary lattice,
\begin{equation}
H_{\rm RF}(t)=-\sum_{\langle ij\rangle }J_{ij}\sigma^z_{i}\sigma^z_{j}-h(t)\sum_{i}\eta _{i}\sigma _{i}^{z},  \label{H1}
\end{equation}
where $\sigma_i^z$ is the $z$ component of Pauli matrix at site $i$ and is considered in this section to be a classical variable taking the values $\pm 1$.
The distribution function of quenched randomness of $J_{ij}$ is specified as 
\begin{eqnarray}
P(J _{ij}) =p\delta (J_{ij}-J)+(1-p)\delta (J_{ij}+J) =\frac{\mathrm{e}%
^{\beta_{p}J_{ij}}}{2\cosh \beta_{p}J}.  \label{distribution}
\end{eqnarray}
The parameter $\beta_p$ has been defined by $\mathrm{e}^{-2\beta_pJ}=(1-p)/p$ to recover the middle expression of eq. (\ref{distribution}) from the right-most expression for $J_{ij}=J$ and $J_{ij}=-J$.
The function $h(t)$ in eq. (\ref{H1}) represents the time-dependent strength of the field starting from $h(0)=h_{0}$ and evolves toward $h(T)=h_{T}$, and $\eta _{i}$ is the quenched random variable following the distribution function
\begin{equation}
P(\eta _{i})=q\delta (\eta _{i}-1)+(1-q)\delta (\eta _{i}+1)=\frac{\mathrm{e}^{\beta _{p}h_{p}\eta _{i}}}{2\cosh \beta _{p}h_{p}}.  \label{distributionh}
\end{equation}
The parameter $h_{p}$ is given by $\exp (-2\beta _{p}h_{p})=(1-q)/q$. 
The following analyses can readily be applied to other distribution functions of $J_{ij}$ and $h_{i}$ as long as they satisfy a certain type of gauge symmetry \cite{HNbook,HN81}.

The above Hamiltonian is invariant under the gauge transformation given by the following simultaneous changes of the signs of interactions,  random fields and spin direction:
\begin{equation}
J_{ij} \to J_{ij}\xi_i\xi_j, ~\eta_i \to \eta_i\xi_i, ~\sigma_i^z \to \sigma_i^z\xi_i\quad(\forall i, j),
\end{equation}
where $\xi_i$ is a gauge variable taking  $\pm 1$.
While the Hamiltonian (\ref{H1}) does not change under the above gauge transformation, the distribution functions (\ref{distribution}) and (\ref{distributionh}) are modified as
\begin{equation}
P(J_{ij}) =\frac{\exp(\beta_p J_{ij}\xi_i\xi_j)}{2\cosh \beta_pJ},\quad
P(\eta_{ij}) =\frac{\exp(\beta_ph_p \eta_{i}\xi_i)}{2\cosh \beta_ph_p}. 
\label{P_gt}
\end{equation}
These properties help us to derive the results in the following sections in addition to the exact value of the internal energy and several exact/rigorous results on a subspace of the phase diagram known as the Nishimori line (NL), $\beta=\beta_p$ and $h(t)=h_p$ \cite{HN81,HNbook}, where $\beta$ is the inverse temperature.

\subsection{Jarzynski equality}

The Jarzynski equality is useful to relate equilibrium and nonequilibrium processes.
The Jarzynski equality in the present problem for a specific configuration of $\{J_{ij}\}$ and $\{\eta_{i}\}$ is written as
\begin{equation}
\langle \mathrm{e}^{-\beta W}\rangle _{\mathrm{RF}}=\frac{Z_{\beta
}(h_{T};\{J_{ij}\},\{\eta _{i}\})}{Z_{\beta }(h_{0};\{J_{ij}\},\{\eta _{i}\})%
},  \label{hJE}
\end{equation}
where $Z_{\beta }(h(t);\{J_{ij}\},\{\eta _{i}\})$ is the partition function of the system (\ref{H1}) (when the system is kept in equilibrium with the Hamiltonian of eq. (\ref{H1})).
The symbol $\langle \cdots \rangle_{\rm RF}$ denotes the average over all possible nonequilibrium processes that start from equilibrium with inverse temperature $\beta$ at $t=0$ and evolve following the master equation. 
The Jarzynski equality holds independently of the details of dynamics and the functional form of $h(t)$. 

We use here a discrete time representation for simplicity, $t=k\Delta t~(k=0,1,2,\cdots)$, where $\Delta t$ is a small time unit.
Correspondingly, the spin configuration changes as $\{\sigma \}_{0}$ ($t=0$), $\{\sigma \}_{1}$ ($t=\Delta t$), $\cdots ,\{\sigma \}_{n}$ ($t=n\Delta t=T$). 
Notice that each $\{\sigma \}_{k}$ stands for an instantaneous classical configuration of $N$ spins as $\{\sigma \}_{k}=\{\sigma _{1}^{z}(t),\sigma _{2}^{z}(t),\cdots,\sigma _{N}^{z}(t)\}$ at time $t=k\Delta t$. 
The performed work in the short interval $\Delta t$ is given by the energy difference due directly to the change of the Hamiltonian 
\begin{equation}
-\delta W_{k}=\{ h((k+1)\Delta t)-h(k\Delta t)\}
\sum_{i}\eta _{i}\sigma _{i}^{z}.
\end{equation}%
The total work $W$ is given by the summation of this infinitesimal work over time, $\sum_{k}\delta W_{k}$.

\subsection{Work performed by random fields}
The Jensen inequality reduces the Jarzynski equality (\ref{hJE}) to the well-known expression of the second law of thermodynamics,
\begin{equation}
\langle W \rangle _{\mathrm{RF}} \ge -\frac{1}{\beta} \log \frac{%
Z_{\beta}(h_T;\{J_{ij}\},\{\eta _{i}\})}{Z_{\beta}(h_0;\{J_{ij}\},\{\eta
_{i}\})}\Big(=F(h_T;\{J_{ij}\},\{\eta _{i}\})-F(h_0;\{J_{ij}\},\{\eta _{i}\})\Big).
  \label{Jen}
\end{equation}
The quantity in the large parentheses is the difference of the equilibrium free energies for the initial and final Hamiltonians. 
It is difficult to obtain the explicit form of the free energy for a specific configuration of $\{ J_{ij}\}$ and $\{\eta_i\}$. 
Therefore we usually evaluate the configurationally-averaged quantity over all realizations of $\{ J_{ij}\}$ and $\{\eta_i\}$. 
Let us thus consider the configurational average of both sides of the above inequality as 
\begin{equation}
\left[\langle W \rangle _{\mathrm{RF}}\right]_{\beta_p,h_p} \ge -\frac{1}{%
\beta} \left[\log \frac{Z_{\beta}(h_T;\{J_{ij}\},\{\eta _{i}\})}{%
Z_{\beta}(h_0;\{J_{ij}\},\{\eta _{i}\})} \right]_{\beta_p,h_p},
\label{Wineq1}
\end{equation}
where the square brackets with the subscript $\beta_p,h_p$ denote the configurational average following the distribution functions of $\{ J_{ij}\}$ and $\{\eta_i\}$, eqs. (\ref{distribution}) and (\ref{distributionh}).
As proved below,  the right-hand side of eq. (\ref{Wineq1}) reduces to, under the NL condition $\beta_p=\beta$ and $h_p=h_0$,
\begin{equation}
\left[\langle W \rangle _{\mathrm{RF}}\right]_{\beta,h_0} \ge \frac{1}{\beta}
D(\beta,h_0|\beta,h_T) -\frac{N}{\beta} \log \left(\frac{\cosh \beta h_T}{%
\cosh \beta h_0}\right),  \label{LowerW}
\end{equation}
where $N$ denotes the number of spins and $D(\beta,h_0|\beta,h_T)$ is the Kullback-Leibler divergence,
\begin{equation}
D(\beta, h_0|\beta,h_T) =
\sum_{\{J_{ij}\},\{\eta_i\}}P_{\beta}(h_0;\{J _{ij}\},\{\eta _{i}\}) \log 
\frac{P_{\beta}(h_0;\{J _{ij}\},\{\eta _{i}\})}{P_{\beta}(h_T;\{J _{ij}\},\{\eta _{i}\})}.
\end{equation}
Here we defined the probability for the configuration of $\{J_{ij}\}, \{\eta_i\}$ summed up over all possible gauge transformations,
\begin{equation}
P_{\beta}(h;\{J _{ij}\},\{\eta _{i}\})
=\frac{1}{2^N}\sum_{\{\xi_i\}}\prod_{\langle ij\rangle}P(J_{ij})\prod_i P(\eta_i)
=\frac{Z_{\beta}(h;\{J _{ij}\},\{\eta
_{i}\})}{2^N(2\cosh \beta J)^{N_B}(2\cosh \beta h)^{N}},\label{g-invariant-prob}
\end{equation}
where $N_B$ expresses the number of bonds, and expressions in eq. (\ref{P_gt})  have been used.
Since the Kullback-Leibler divergence is non-negative, the work performed by random fields during a nonequilibrium process from the NL condition (i.e. the left-hand side of eq. (\ref{LowerW})) does not exceed the second quantity on the right-hand side of eq. (\ref{LowerW}),
\begin{equation}
\left[\langle W \rangle _{\mathrm{RF}}\right]_{\beta,h_0} \ge -\frac{N}{%
\beta} \log \left(\frac{\cosh \beta h_T}{\cosh \beta h_0}\right).
\label{LBL0}
\end{equation}

To prove eq. (\ref{LowerW}), we apply the gauge transformation as 
\begin{eqnarray}
& & \left[\log \frac{Z_{\beta}(h_T;\{J _{ij}\},\{\eta _{i}\})}{%
Z_{\beta}(h_0;\{J _{ij}\},\{\eta _{i}\})} \right]_{\beta_p,h_p}  \nonumber \\
& & = \sum_{\{J_{ij}\},\{\eta_i\}}\frac{\exp\big(\beta_p\sum_{\langle ij
\rangle}J_{ij}\xi_i\xi_j+\beta_ph_p\sum_i\eta_i\xi_i\big)}{(2\cosh
\beta_pJ)^{N_B}(2\cosh \beta_ph_p)^{N}}\log \frac{Z_{\beta}(h_T;\{J
_{ij}\},\{\eta _{i}\})}{Z_{\beta}(h_0;\{J _{ij}\},\{\eta _{i}\})}.
\end{eqnarray}
Then we sum it up over all possible configurations of $\{\xi_i\}$ and divide the result by $2^N$ to obtain 
\begin{eqnarray}
& & \left[\log \frac{Z_{\beta}(h_T;\{J _{ij}\},\{\eta _{i}\})}{%
Z_{\beta}(h_0;\{J _{ij}\},\{\eta _{i}\})} \right]_{\beta_p,h_p}  \nonumber \\
& & = \sum_{\{J_{ij}\},\{\eta_i\}}\frac{Z_{\beta_p}(h_p;\{J_{ij},\{\eta_i\}%
\})}{2^N(2\cosh \beta_pJ)^{N_B}(2\cosh \beta_ph_p)^{N}}\log \frac{%
Z_{\beta}(h_T;\{J _{ij}\},\{\eta _{i}\})}{Z_{\beta}(h_0;\{J _{ij}\},\{\eta
_{i}\})}.
\end{eqnarray}
According to eq. (\ref{g-invariant-prob}), this leads to eq. (\ref{LowerW}) under the NL condition $\beta_p=\beta$ and $h_p = h_0$.

Instead of the inequality (\ref{Jen}), we may consider the configurational average of eq. (\ref{hJE}) itself. 
Then we obtain the following simple equation by similar calculations as in the above analysis, 
\begin{equation}
\left[\langle \mathrm{e}^{-\beta W} \rangle _{\mathrm{RF}}\right]%
_{\beta,h_0} = \left(\frac{\cosh \beta h_T}{\cosh \beta h_0}\right)^{N}.
\label{JESGh}
\end{equation}
Again, by the Jensen inequality, a lower bound of the performed work is evaluated, leading to the same inequality as eq. (\ref{LBL0}),
\begin{equation}
\left[\langle W \rangle _{\mathrm{RF}}\right]_{\beta,h_0} \ge -\frac{N}{%
\beta} \log \left(\frac{\cosh \beta h_T}{\cosh \beta h_0}\right).
\label{LBL}
\end{equation}
This is a bound looser than the previous one, eq. (\ref{LowerW}), since the Kullback-Leibler divergence does not appear here.
We emphasize that eq. (\ref{JESGh}) is highly non-trivial because it gives an explicit bound for the work performed on spin glasses in nonequilibrium processes.

\subsection{Relation with different quenched randomness}\label{section:Relationship}
By the method used in the previous study on the Jarzynski equality in spin glasses without external fields\cite{ON}, we can derive several exact relations between equilibrium and nonequilibrium quantities for spin glasses in the presence of random longitudinal fields.
The results in this section are relatively straightforward generalizations of those in our previous paper\cite{ON}, and therefore detailed derivations are omitted here.

It is not difficult to show an identity for an observable $O_{T}$ measured at the final step of time evolution, using the fluctuation theorem,
\begin{equation}
\left\langle O_{T}\mathrm{e}^{-\beta W}\right\rangle _{\rm RF}=\langle O\rangle _{\beta,h_T}\frac{Z_{\beta}(h_{T};\{J_{ij}\},\{\eta_{i}\})}{Z_{\beta}(h_{0};\{J_{ij}\},\{\eta_{i}\})},  \label{JE3}
\end{equation}
where $\langle O \rangle_{\beta,h_T}$ denotes the equilibrium (thermal) average of the observable $O$ in the final equilibrium state with the strength of the random field $h_T$.
This relation leads us to several unexpected equalities in spin glasses.
Calculations by use of the gauge symmetry, similar to the above one, give the following identities
\begin{eqnarray}
\left[ \left\langle H_{\rm RF}(T)\mathrm{e}^{-\beta W}\right\rangle _{\rm RF}\right]_{\beta,h_0} = \left[ \langle H_{\rm RF} \rangle _{\beta,h_{T}}\right]
_{\beta,h_{T}}\left(\frac{\cosh \beta h_T}{\cosh \beta h_0}\right)^{N}.  \label{Ene}
\end{eqnarray}
We remark that different concentrations of quenched randomness, represented by $h_0$ and $h_T$, appear in both sides on the equality.
The same relation holds for other gauge invariant quantities.

In addition to the gauge-invariant quantities, the gauge transformation yields exact equalities on the local magnetization $\sigma_i^z$ and correlation function $\sigma_0^z \sigma_r^z$, both of which are gauge non-invariant, as
\begin{equation}
\left[ \langle \sigma_{i}^z(T)\mathrm{e}^{-\beta W}\rangle _{\rm RF}\right] _{\beta,h_{0}}=\left[ \langle \sigma_{i}^z\rangle _{\beta,h_{0}}\right] _{\beta,h_{T}}\left( 
\frac{\cosh \beta h_{T}}{\cosh \beta h_{0}}\right) ^{N} \label{BJE1}
\end{equation}
and
\begin{eqnarray}
\left[ \langle \sigma_{0}^z(T)\sigma_{r}^z(T)\mathrm{e}^{-\beta W}\rangle
_{\rm RF}\right] _{\beta,h_{0}} =\left[ \langle \sigma_{0}^z\sigma_{r}^z\rangle
_{\beta,h_0}\right] _{\beta,h_{T}}\left( \frac{\cosh \beta h_{T}}{\cosh \beta h_{0}}\right)
^{N}.  \label{BJE2}
\end{eqnarray}
These results relate physical quantities measured in quite different environments.
They imply the possibility that equilibrium physical quantities in spin glasses, the right-hand sides of eqs. (\ref{BJE1}) and (\ref{BJE2}), can be evaluated from nonequilibrium calculations (left-hand sides) in different parts of the phase diagram with the aid of annealed importance sampling or population annealing \cite{NJ,Iba,Pop1}.

Furthermore, the fluctuation theorem \cite{J3,J4,J5} yields the following relation for a quantity that depends on the intermediate spin configurations,
 $O(\{\sigma\}_0,\{\sigma\}_1,\cdots,\{\sigma\}_T)$, as
\begin{equation}
\left\langle O(\{\sigma\}_0,\{\sigma\}_1,\cdots,\{\sigma\}_T)\mathrm{e}^{-\beta W}\right\rangle _{\rm RF}=\langle O_{\rm r}(\{\sigma\}_0,\{\sigma\}_1,\cdots,\{\sigma\}_T)\rangle^{\rm rev.}_{\rm RF}\frac{Z_{\beta}(h_{T};\{J_{ij}\},\{\eta_{i}\})}{Z_{\beta}(h_{0};\{J_{ij}\},\{\eta_{i}\})},  \label{JE3}
\end{equation}
where $O_{\mathrm{r}}$ denotes the observable that depends on the backward process $h_{T}\rightarrow h_{0}$, and the angular brackets with the superscript `rev.' express the nonequilibrium average for the backward process. 
An application of the gauge transformation to eq. (\ref{JE3}) gives the following exact equality for the autocorrelation function $O(\{\sigma\}_0,\{\sigma\}_1,\cdots,\{\sigma\}_T)=\sigma_{i}^z(0)\sigma_{i}^z(T)$,
\begin{eqnarray}
\left[ \left\langle \sigma_{i}^z(0)\sigma_{i}^z(T)\mathrm{e}^{-\beta W}\right\rangle
_{\rm RF}\right] _{\beta,h_{0}} =\left[ \langle \sigma_{i}^z(0)\sigma_{i}^z(T)\rangle^{\rm rev.} _{\rm RF}\right] _{\beta,h_{T}}\left( 
\frac{\cosh \beta h_{T}}{\cosh \beta h_{0}}\right) ^{N}.  \label{NE1}
\end{eqnarray}
This result states that the equilibrium autocorrelation function in the reversed process can be computed from the nonequilibrium process in a different part of the phase diagram.

\section{Jarzynski equality for quantum annealing}
\subsection{Spin glass in transverse field}
Let us next consider another system with a transverse field,
\begin{equation}
H_{\rm QA}(t)=-g(t)\sum_{\langle ij\rangle }J_{ij}\sigma^z_{i}\sigma^z_{j} -\left( 1-g(t)\right) \Gamma _{0}\sum_{i}\sigma _{i}^{x},
\label{H2}
\end{equation}
where $g(t)=t/T$, which changes from 0 to 1 as $t$ goes from 0 to $T$.
This system is used in quantum annealing for search of the ground state of the spin glass Hamiltonian\cite{QA1,QA2,QA3}
\begin{equation}
 H_0=-\sum_{\langle ij\rangle }J_{ij}\sigma^z_{i}\sigma^z_{j},
\end{equation}
which is shared with eq. (\ref{H1}).
The whole Hamiltonian is invariant under the gauge transformation,
\begin{equation}
\sigma_i^x\to \sigma_i^x,~\sigma_i^y\to \xi_i\sigma_i^y,~\sigma_i^z\to \xi_i\sigma_i^z,~J_{ij} \to J_{ij}\xi_i\xi_j\quad (\forall i,j),
\end{equation}
where $\xi_i (=\pm 1)$ is a gauge variable.
Notice that this transformation is designed to preserve the commutation relations between different components of Pauli matrix \cite{Morita}.

The adiabatic theorem guarantees that a sufficiently slow decrease of the strength of the transverse field (i.e. large $T$) changes the trivial initial state, the ground state of $-\Gamma_0\sum_i\sigma_i^x$, to the nontrivial ground state of the target Hamiltonian $H_{0}$. 
This is a special case of quantum annealing, quantum adiabatic computation\cite{QAA}.
Quantum adiabatic computation, however, is known to be unable to solve efficiently certain instances of hard optimization problems \cite{FT1,FT2}. 
Thus, instead of the adiabatic control, we analyze a method to repeat non-adiabatic quantum annealing (small or intermediate $T$) starting from a state chosen from equilibrium ensemble, not necessary the ground state.
We may not be able to easily reach the ground state of $H_0$ by such processes even if we start from a very low-temperature state since the system does not trace the instantaneous ground state as in the adiabatic evolution. 
We instead need to repeat the process many times to accurately evaluate the average of the exponentiated work over non-adiabatic processes appearing in the Jarzynski equality.
In this way, the problem of long annealing time is replaced by a problem of very many repetitions of non-adiabatic (possibly quick) evolution.
We analyze such non-adiabatic quantum annealing using the Jarzynski equality and gauge symmetry.

\subsection{Non-adiabatic quantum annealing}
Initially we pick up a state from the canonical ensemble for $H_{\rm QA}(0)=-\Gamma_0\sum_i\sigma_i^x$ and then let it evolve following the time-dependent Schr\"{o}dinger equation.
The performed work in the present quantum problem is given by the difference between the outputs of projective measurements of the initial and final energies, $W=E_{m}(T)-E_{n}(0)$. 
Here $m$ and $n$ denote the indices of the instantaneous eigenstates measured at the final and initial steps, $H_{\rm QA}(T)|m(T)\rangle =E_{m}(T)|m(T)\rangle $ and $H_{\rm QA}(0)|n(0)\rangle=E_{n}(0)|n(0)\rangle $, respectively.
The Jarzynski equality is \cite{QJE1,QJE2}
\begin{equation}
\langle \mathrm{e}^{-\beta W}\rangle _{\mathrm{QA}}=\frac{Z_{\beta
}(T,\{J_{ij}\})}{Z_{\beta }(0;\{J_{ij}\})},  \label{qJE}
\end{equation}%
where $Z(t;\{J_{ij}\})$ is the partition function for the instantaneous Hamiltonian (\ref{H2}). 
The left-hand side of eq. (\ref{qJE}) expresses the average of the exponentiated work over all realizations of non-adiabatic processes starting from the equilibrium ensemble.

Following the prescription of the Jarzynski equality, we consider a repetition of non-adiabatic quantum annealing starting from the equilibrium ensemble.
The initial Hamiltonian is given only by the transverse field, which means a trivial initial distribution. 
Consequently, the Jarzynski equality (\ref{qJE}) for non-adiabatic quantum annealing reduces to 
\begin{equation}
\langle \mathrm{e}^{-\beta W}\rangle _{\mathrm{QA}}=\frac{Z_{\beta
}(T,\{J_{ij}\})}{(2\cosh \beta \Gamma _{0})^{N}}.  \label{QAJE}
\end{equation}%

\subsection{Work performed by the transverse field}
\begin{figure}[tb]
\begin{center}
\includegraphics[width=80mm]{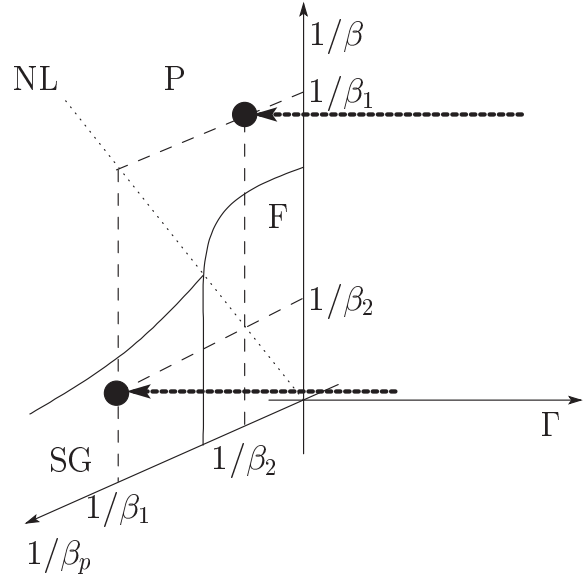}
\end{center}
\caption{{\protect\small Two processes of non-adiabatic quantum annealing in eq. (\ref{QAdiff}). 
The left-hand side of eq. (\ref{QAdiff}) corresponds to the annealing process ending at the upper-right black dot and the right-hand side terminates at the lower-left dot.
Three phases (F: Ferromagnetic, P: paramagnetic, and SG: Spin Glass) are separated by solid curves and a vertical line.
The dotted line expresses the NL.
}}
\label{fig1}
\end{figure}

Let us take the configurational average of eq. (\ref{QAJE}) over all realizations of $\{J_{ij}\}$ for $\beta=\beta_1$ and $\beta_p=\beta_2$ as 
\begin{equation}
\left[ \langle \mathrm{e}^{-\beta_1 W}\rangle _{\mathrm{QA}}\right] _{\beta
_2}=\left[ \frac{Z_{\beta_1 }(T;\{J_{ij}\})}{\left( 2\cosh \beta_1 \Gamma
_{0}\right) ^{N}}\right] _{\beta _2}.
\end{equation}%
The right-hand side is written explicitly as
\begin{equation}
\left[ \langle \mathrm{e}^{-\beta_1 W}\rangle _{\mathrm{QA}}\right] _{\beta_2}=\sum_{\{J_{ij}\}}\frac{\exp \big(\beta_2\sum_{\langle ij\rangle }J_{ij}\big)%
}{(2\cosh \beta_2 J)^{N_{B}}}\frac{Z_{\beta_1 }(T;\{J_{ij}\})}{\left( 2\cosh\beta_1 \Gamma _{0}\right) ^{N}}.
\end{equation}%
Let us apply the gauge transformation and sum over all possible configurations of the gauge variables $\{\xi _{i}\}$. 
We then obtain, after dividing the result by $2^{N}$, 
\begin{equation}
\left[ \langle \mathrm{e}^{-\beta_1 W}\rangle _{\mathrm{QA}}\right] _{\beta_2}=\sum_{\{J_{ij}\}}\frac{Z_{\beta_2}(T;\{J_{ij}\})Z_{\beta_1}(T;\{J_{ij}\})}{2^{N}(2\cosh \beta_2J)^{N_{B}}\left( 2\cosh \beta_1 \Gamma
_{0}\right) ^{N}}.  \label{QA1}
\end{equation}%
A similar average of the exponentiated work on spin glass with the inverse temperature $\beta _2$ and the parameter for the quenched randomness $\beta_1 $ gives
\begin{equation}
\left[ \langle \mathrm{e}^{-\beta_2W}\rangle _{\mathrm{QA}}\right]
_{\beta_1 }=\sum_{\{J_{ij}\}}\frac{Z_{\beta_2 }(T;\{J_{ij}\})Z_{\beta_1}(T;\{J_{ij}\})}{2^{N}(2\cosh \beta_1 J)^{N_{B}}\left( 2\cosh \beta_2 \Gamma _{0}\right) ^{N}}.  \label{QA2}
\end{equation}%
Comparing eqs. (\ref{QA1}) and (\ref{QA2}), we find the following relation between different non-adiabatic processes,
\begin{equation}
\left[ \langle \mathrm{e}^{-\beta_1 W}\rangle _{\mathrm{QA}}\right] _{\beta_2}=\left[ \langle \mathrm{e}^{-\beta_2 W}\rangle _{\mathrm{QA}}\right]
_{\beta_1}\left( \frac{\cosh \beta_1 J}{\cosh \beta_2 J}\right)
^{N_{B}}\left( \frac{\cosh \beta_2 \Gamma _{0}}{\cosh \beta_1 \Gamma _{0}}%
\right) ^{N}.\label{QAdiff}
\end{equation}
Figure \ref{fig1} describes the two different paths of non-adiabatic quantum annealing related by this equality.
If we set $\beta_2=0$ in eq. (\ref{QAdiff}), (implying $p=1/2$, the symmetric distribution or the high-temperature limit), we obtain a simple equality on the performed work during non-adiabatic quantum annealing 
\begin{equation}
\left[ \langle \mathrm{e}^{-\beta_1 W}\rangle _{\mathrm{QA}}\right] _{0}=\frac{(\cosh \beta_1 J)^{N_{B}}}{(\cosh \beta_1 \Gamma _{0})^{N}}.
\label{QA01}
\end{equation}
The symmetric distribution ($\beta_2=0$ on the left-hand side) makes it possible to reduce the right-hand side to the above trivial expression.
It is remarkable that non-adiabatic quantum annealing, which involves very complex dynamics, satisfies such a simple identity irrespective of the speed of annealing $T$.

Similarly to the classical case discussed in \S \ref{section:Relationship}, we can formulate a relation between equilibrium and nonequilibrium quantities as
\begin{equation}
\langle O_T\mathrm{e}^{-\beta W}\rangle _{\mathrm{QA}}=\langle O \rangle_{\beta} \frac{Z_{\beta}(T,\{J_{ij}\})}{(2\cosh \beta \Gamma _{0})^{N}},  \label{QAJE2}
\end{equation}
where $O_T$ is an observable measured at the final time.
The angular brackets with subscript $\beta$ is the equilibrium (thermal) average by the final Hamiltonian $H_{\rm QA}(T)=H_0$.
If we choose $H_0$ as the observable $O$ and take the configurational average with the symmetric distribution for both sides of the above equality, we obtain 
\begin{equation}
\left[\langle H_{\rm QA}(T)\mathrm{e}^{-\beta W}\rangle_{\mathrm{QA}}\right]_{0}= -\frac{JN_B(\cosh \beta J)^{N_B}}{(\cosh \beta \Gamma_0)^{N}}\tanh\beta J.\label{HQA}
\end{equation}
The quantity on the right-hand side is obtained with the relationship given by the gauge transformation as
\begin{equation}
\left[\langle O \rangle_{\beta} \frac{Z_{\beta}(T,\{J_{ij}\})}{(2\cosh \beta \Gamma _{0})^{N}}\right]_{0} = \frac{(\cosh \beta J)^{N_B}}{(\cosh \beta \Gamma _{0})^{N}}\left[\langle O \rangle_{\beta}\right]_{\beta}.
\end{equation}
The identity $[\langle H_0\rangle_{\beta}]_{\beta}=-JN_B\tanh \beta J$ on the NL \cite{HN81,HNbook} leads us to eq. (\ref{HQA}).
Equation (\ref{HQA}) reads
\begin{equation}
\frac{\left[\langle H_{\rm QA}(T)\mathrm{e}^{-\beta W}\rangle_{\mathrm{QA}}\right]_{0}}{\left[\langle\mathrm{e}^{-\beta W}\rangle_{\mathrm{QA}}\right]_{0}}= -N_B J\tanh\beta J.
\end{equation}
Similarly to the classical case, we have obtained the equilibrium quantity (right-hand side) defined away from the part of the phase diagram where the original non-adiabatic process is performed (left-hand side).

\subsection{Exact relations involving inverse statistics}
\begin{figure}[tb]
\begin{center}
\includegraphics[width=70mm]{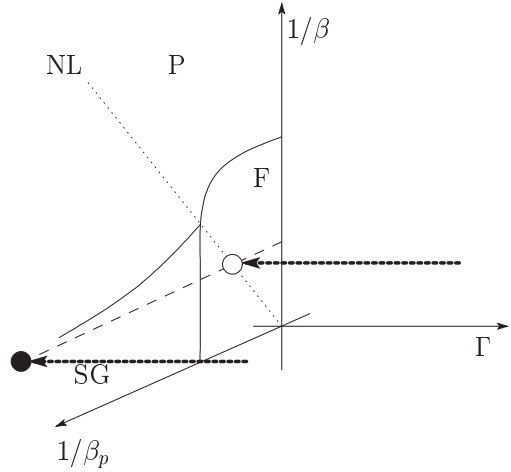}
\end{center}
\caption{{\protect\small Two different nonequilibrium processes of non-adiabatic quantum annealing are related through  eq. (\ref{IQAdiff}).
The same symbols are used as in Fig. \ref{fig1}.
The white circle denotes the target of the process on
the right-hand side of eq. (\ref{IQAdiff}), whereas the black dot is for the left-hand side.
}}
\label{fig2}
\end{figure}

Let us next take the configurational average of the inverse of the Jarzynski equality, eq. (\ref{QAJE}), as 
\begin{equation}
\left[\frac{1}{\langle \mathrm{e}^{-\beta W} \rangle_{\mathrm{QA}}}\right]%
_{\beta_p} = \left[\frac{\left( 2 \cosh \beta\Gamma_0 \right)^{N}}{Z_{\beta}(T;\{J_{ij}\})}\right]_{\beta_p}
\label{QAinverse0}
\end{equation}
When the gauge transformation is applied to the right-hand side, we have
\begin{equation}
\left[\frac{1}{\langle \mathrm{e}^{-\beta W} \rangle_{\mathrm{QA}}}\right]%
_{\beta_p} = \sum_{\{J_{ij}\}}\frac{\exp\big(\beta_p\sum_{\langle ij
\rangle}J_{ij}\xi_i\xi_j\big)}{(2\cosh \beta_p J)^{N_B}}\frac{\left( 2 \cosh
\beta\Gamma_0 \right)^{N}}{Z_{\beta}(T;\{J_{ij}\})}.
\end{equation}
By summing the right-hand side over all possible configurations of $\{\xi_i\}$ and dividing the result by $2^N$, we find
\begin{equation}
\left[\frac{1}{\langle \mathrm{e}^{-\beta W} \rangle_{\mathrm{QA}}}\right]%
_{\beta_p} = \sum_{\{J_{ij}\}}\frac{Z_{\beta_p}(T;\{J_{ij}\})}{2^N(2\cosh
\beta_p J)^{N_B}}\frac{\left( 2 \cosh \beta\Gamma_0 \right)^{N}}{%
Z_{\beta}(T;\{J_{ij}\})}.  \label{IQA}
\end{equation}
Under the NL condition $\beta_p = \beta$, this equation reduces to
\begin{equation}
\left[\frac{1}{\langle \mathrm{e}^{-\beta W} \rangle_{\mathrm{QA}}}\right]%
_{\beta} = \frac{(\cosh \beta\Gamma_0)^N}{(\cosh \beta J)^{N_B}}.\label{QA02}
\end{equation}
Comparison of eqs. (\ref{QA01}) and (\ref{QA02}) reveals
\begin{equation}
\left[\langle \mathrm{e}^{-\beta W} \rangle_{\mathrm{QA}}\right]_{0} = \left(%
\left[\frac{1}{\langle \mathrm{e}^{-\beta W} \rangle_{\mathrm{QA}}}\right]%
_{\beta}\right)^{-1}.\label{IQAdiff}
\end{equation}
As depicted in Fig. \ref{fig2}, two completely different processes are related by this equation:
One toward the NL and the other for the symmetric distribution.

Let us further consider the inverse of eq. (\ref{QAJE2}) for $O_T = \sigma^z_i\sigma^z_j$.
We take the configurational average of both sides under the NL condition as
\begin{equation}
\left[\frac{1}{\langle \sigma^z_i\sigma^z_j\mathrm{e}^{-\beta W} \rangle_{\mathrm{QA}}}\right]%
_{\beta} = \frac{(\cosh \beta\Gamma_0)^N}{(\cosh \beta J)^{N_B}}\left[\frac{1}{\langle\sigma^z_i\sigma^z_j\rangle_{\beta}}\right]_{\beta}.
\end{equation}
The quantity on the right-hand side is unity by the gauge transformation as has been shown in the literatures \cite{HN81,HNbook}.
We thus obtain a simple exact relation
\begin{equation}
\left[\frac{1}{\langle \sigma^z_i\sigma^z_j\mathrm{e}^{-\beta W} \rangle_{\mathrm{QA}}}\right]%
_{\beta} = \frac{(\cosh \beta\Gamma_0)^N}{(\cosh \beta J)^{N_B}},
\end{equation}
which is another exact identity for processes of non-adiabatic quantum annealing.

\subsection{Possibility as a solver}

The last part of this section is devoted to a discussion on the possibility to measure equilibrium quantities through non-adiabatic quantum annealing.
The ratio of eqs. (\ref{QAJE}) and (\ref{QAJE2}) gives
\begin{equation}
\frac{\langle \hat{O}_T\mathrm{e}^{-\beta W}\rangle_{\mathrm{QA}}}{\langle\mathrm{e}^{-\beta W}\rangle _{\mathrm{QA}}} =\langle \hat{O} \rangle_{\beta}.  \label{QAJE3}
\end{equation}
This equation suggests that the equilibrium (canonical) average under the Hamiltonian $H_0$ on the right-hand side can be estimated by the non-adiabatic quantum annealing on the left-hand side.
This fact may be useful in the evaluation of equilibrium average when it takes a very long time to equilibrate the system in Monte Carlo simulations as in spin glasses, since the left-hand side is evaluated without slow adiabatic processes.
Nevertheless we should be careful because the average on the left-hand side involves a non-extensive quantity, the exponentiated work, whose value fluctuates significanlty from process to process:
Remember that the average on the left-hand side is calculated by many trials of annealing processes.
Thus, rare events with large values of the exponentiated work (i.e. $\beta |W|\gg 1$) would contribute to the average significantly, and we have to repeat the annealing process very many times in order to reach the correct value of the average.

The low-temperature limit of the above argument gives us a suggestion for non-adiabatic quantum annealing to identify the ground state.
If the temperature is lower than the energy gap between the ground state and the first excited state of $H_0$ and $\beta \gg \Gamma_0$, the Jarzynski equality (\ref{qJE}) is reduced to
\begin{equation}
\langle \mathrm{e}^{-\beta W}\rangle _{\mathrm{QA}}\approx \exp (-\beta E_{\mathrm{%
GS}}(\{J_{ij}\})+\beta N\Gamma _{0}),  \label{NQA}
\end{equation}
where $E_{\rm GS}(\{ J_{ij}\})$ is the ground-state energy of $H_0$.
This equation suggests that the ground-state energy can be evaluated by the repetition of non-adiabatic quantum annealing, which implies that we would hit the correct ground state if we repeat the non-adiabatic process many times.
We have thus replaced the problem of long annealing time in quantum adiabatic computation by another problem of many repetitions of non-adiabatic (fast) quantum annealing.
It usually takes very many, typically exponentially many, repetitions to correctly evaluate the left-hand side of eq. (\ref{NQA}).
Thus the difficulty has not been relaxed in general, but the present new perspective may lead to different methods and tools than conventional ones to attack the problem.

A related remark is that, for classical systems, an improvement by a modification of the dynamics has succeeded in estimating the free energy difference through fast nonequilibrium processes through the Jarzynski equality \cite{Escorted}.
A similar idea has been realized for quantum annealing using another degree of freedom \cite{Ohzeki}.

\section{Simulated annealing and quantum annealing}
It is possible to establish identities to relate completely different annealing processes.
The Jarzynski equality holds for simulated annealing of the classical system $H_0$ through the `pseudo work' defined by
\begin{equation}
-\delta Y(k\Delta t)=-\left( \beta _{(k+1)\Delta t}-\beta _{k\Delta
t}\right) H_{0},
\end{equation}%
where the inverse temperature is changed from $\beta _{k\Delta t}$ to $\beta _{(k+1)\Delta t}$ at the $k$th step.
Then the Jarzynski equality is
\begin{equation}
\langle \mathrm{e}^{-Y(\beta _{0}\rightarrow \beta _{T})}\rangle _{\mathrm{SA%
}}=\frac{Z_{\beta _{T}}(T;\{J_{ij}\})}{Z_{\beta _{0}}(T;\{J_{ij}\})},
\label{SAJE}
\end{equation}%
where $Y(\beta _{0}\rightarrow \beta _{T})$ is  the sum $\sum_k \delta Y(k\Delta t)$. 
Using the pseudo work $Y$ in the weight of importance sampling, population annealing is implemented to produce the equilibrium ensemble \cite{Iba,Pop1}.

\subsection{Identities for simulated annealing}
Let us consider the configurational average of eq. (\ref{SAJE}) over quenched randomness of $\{J_{ij}\}$ in $H_0$ with $\beta_p = \beta_r$, where $\beta_r$ is a given constant,
\begin{equation}
\left[\langle \mathrm{e}^{-Y(\beta_0 \to \beta_T)} \rangle_{\mathrm{SA}}\right]%
_{\beta_r} = \sum_{\{J_{ij}\}}\frac{\exp\big(\beta_r\sum_{\langle ij
\rangle}J_{ij}\big)}{(2\cosh \beta_r J)^{N_B}}\frac{Z_{\beta_T}(T;\{J_{ij}\})}{%
Z_{\beta_0}(T;\{J_{ij}\})}.
\end{equation}
We apply the gauge transformation and sum over all configurations of $%
\{\xi_i\}$. Division by $2^N$ of the result leads to 
\begin{equation}
\left[\langle \mathrm{e}^{-Y(\beta_0 \to \beta_T)} \rangle_{\mathrm{SA}}%
\right]_{\beta_r} = \sum_{\{J_{ij}\}}\frac{Z_{\beta_r}(T;\{J_{ij}\})}{%
2^N(2\cosh \beta_r J)^{N_B}}\frac{Z_{\beta_T}(T;\{J_{ij}\})}{%
Z_{\beta_0}(T;\{J_{ij}\})}.  \label{SA1}
\end{equation}
It is useful to write a similar equation for the case starting
from  $\beta_0$ and ending at $\beta_r$
with the quenched randomness characterized by $\beta_T$, 
\begin{equation}
\left[\langle \mathrm{e}^{-Y(\beta_0 \to \beta_r)} \rangle_{\mathrm{SA}}%
\right]_{\beta_T} = \sum_{\{J_{ij}\}}\frac{Z_{\beta_T}(T;\{J_{ij}\})}{%
2^N(2\cosh \beta_T J)^{N_B}}\frac{Z_{\beta_r}(T;\{J_{ij}\})}{%
Z_{\beta_0}(T;\{J_{ij}\})}.  \label{SA2}
\end{equation}
Equations (\ref{SA1}) and (\ref{SA2}) relates two different annealing processes 
\begin{figure}[tb]
\begin{center}
\includegraphics[width=70mm]{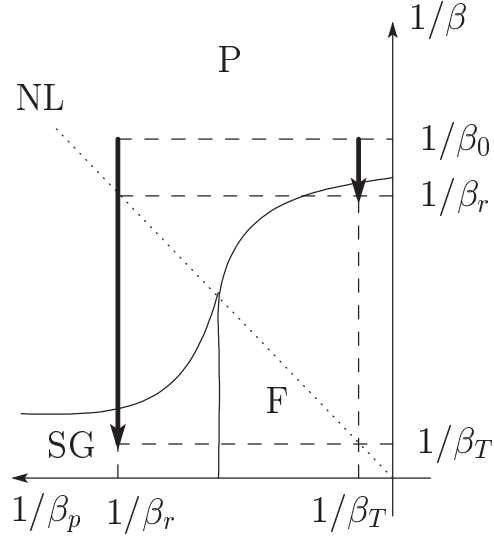}
\end{center}
\caption{{\protect\small Two related nonequilibrium processes of simulated annealing in eq. (\ref{SAdiff}).
The same abbreviations and symbols are used as in Fig. \ref{fig1}.
}}
\label{fig3}
\end{figure}
\begin{equation}
\left[\langle \mathrm{e}^{-Y(\beta_0 \to \beta_T)} \rangle_{\mathrm{SA}}%
\right]_{\beta_r} = \left( \frac{\cosh \beta_T J}{\cosh \beta_r J}%
\right)^{N_B} \left[\langle \mathrm{e}^{-Y_{\mathrm{SA}}(\beta_0 \to
\beta_r)} \rangle_{\mathrm{SA}}\right]_{\beta_T},\label{SAdiff}
\end{equation}
as depicted in Fig. \ref{fig3}.
In particular, if we set $\beta_0=\beta_r$, we can rederive one of the equalities given in
our previous study \cite{ON} as 
\begin{equation}
\left[\langle \mathrm{e}^{-Y(\beta_0 \to \beta_T)} \rangle_{\mathrm{SA}}%
\right]_{\beta_0} = \left( \frac{\cosh \beta_T J}{\cosh \beta_0 J}%
\right)^{N_B}.
\end{equation}

\subsection{Quantum annealing and simulated annealing}
\begin{figure}[tb]
\begin{center}
\includegraphics[width=70mm]{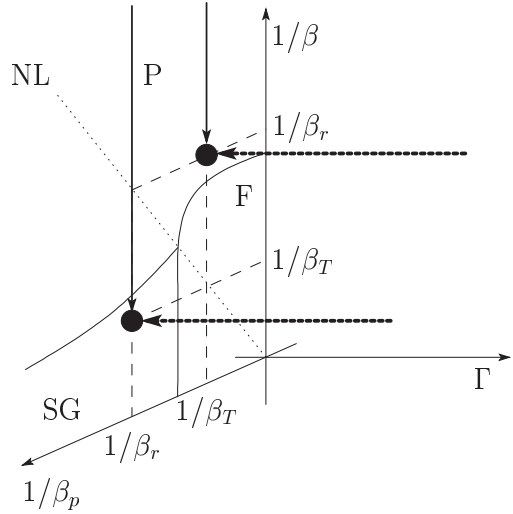}
\end{center}
\caption{{\protect\small Several relations between simulated annealing and quantum annealing by eqs. (\ref{QAdiff}), (\ref{SAdiff}), and (\ref{QASA}).
The same abbreviations and symbols are used as in Fig. \ref{fig1}.
}}
\label{fig4}
\end{figure}

Collecting the above results, we can derive several relations between simulated annealing and quantum annealing. 
We combine eq. (\ref{QA1}) with eq. (\ref{SA1}) for the case starting from $\beta_0 = 0$ and ending at $\beta_T= \beta_1$ with the quenched randomness satisfying $\beta_r=\beta_2$ to obtain 
\begin{equation}
\left[\langle \mathrm{e}^{-\beta_T W} \rangle_{\mathrm{QA}}\right]_{\beta_r} =
\left( \frac{1}{\cosh \beta_T \Gamma_0}\right)^{N} \left[\langle \mathrm{e}%
^{-Y(\beta_0 \to \beta_T)} \rangle_{\mathrm{SA}}\right]_{\beta_r}. \label{QASA}
\end{equation}
This equation relates two processes ending at the black dot in the left-lower part of Fig. \ref{fig4}.
Furthermore, the combination of eqs. (\ref{QAdiff}), (\ref{SAdiff}) and (\ref{QASA}) reveals a close relation between four annealing processes, classical and quantum, drawn in arrows in Fig. \ref{fig4}.

\begin{figure}[tb]
\begin{center}
\includegraphics[width=70mm]{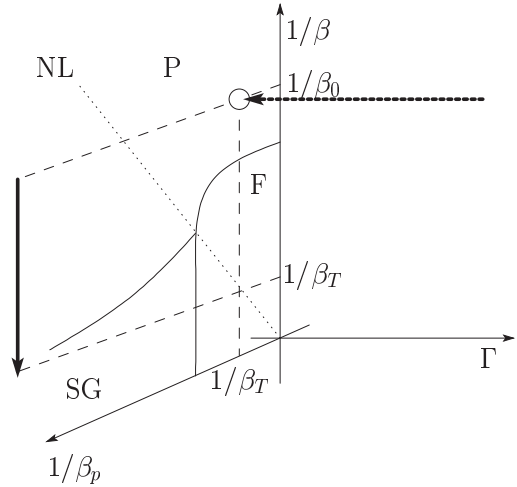}
\end{center}
\caption{{\protect\small The relation between simulated annealing and quantum annealing in eq. (\ref{SAIQA}) is shown in two arrows.
}}
\label{fig5}
\end{figure}

In addition, a comparison of eq. (\ref{IQA}) with eq. (\ref{SA1}) for $\beta = \beta_0$, $\beta_p = \beta_T$ and $\beta_r =0$ shows
\begin{equation}
\left[\frac{1}{\langle \mathrm{e}^{-\beta_0 W} \rangle_{\mathrm{QA}}}\right]%
_{\beta_T} = \frac{\left( \cosh \beta_0\Gamma \right)^{N}}{\left( \cosh \beta_T J \right)^{N_B}}\left[\langle \mathrm{e}^{-Y(\beta_0 \to \beta_T)}
\rangle_{\mathrm{SA}}\right]_{0}.\label{SAIQA}
\end{equation}
Figure \ref{fig5} illustrates this relation between the inverse statistics of the performed work in quantum annealing and the exponentiated pseudo work in simulated annealing.

The identities derived above constitute a class of new exact relations between classical and quantum systems.

\section{Summary}
We have established a number of exact identities and inequalities for nonequilibrium (dynamical) properties of spin glasses in external fields.
Essential ingredients in the analysis were the gauge symmetry of the Hamiltonian and the Jarzynski equality (or the fluctuation theorem).
The results are to be contrasted with our previous study where we discussed spin glass systems in the absence of external field and derived identities involving pseudo work for the change of the temperature \cite{ON}.
The present paper deals with genuine work realized by the change of the strengths of external fields.
The identities and inequalities derived here are rare examples of exact results on nonequilibrium properties of spin glass systems and are expected to lay a foundation of further analyses of dynamical properties of spin glasses and related problems.

The Jarzynski equality in the case of transverse field would be a step toward non-adiabatic realization of quantum annealing.
We remark that it is necessary to repeat the nonequilibrium processes exponentially many times in general to correctly evaluate the nonequilibrium average in eq. (\ref{NQA}).
In this sense we have not solved the problem of exponentially long computation time for hard optimization problems
in the usual quantum adiabatic evolution but have replaced it with exponentially many trials
of quick (non-adiabatic) processes.
We nevertheless expect that the present new point of view would provide a perspective different from previous studies, which may help us approach the problem by different methods, eventually leading to unexpected results.

Also, the non-trivial relations between the exponentiated work in quantum and simulated annealing may become a valuable tool to compare performance of these two generic algorithms for optimization problems.

\begin{acknowledgments}
This work was partially supported by CREST, JST, and by the 21st Century Global COE Program at Tokyo Institute of Technology `Nanoscience and Quantum Physics'.
\end{acknowledgments}

\end{document}